\documentclass[aps,pre,print,twocolumn]{revtex4-1}
\usepackage[utf8]{inputenc}
\usepackage{graphicx}
\usepackage{float}
\usepackage{amsmath}
\usepackage{xcolor,soul}
\sethlcolor{red}
\usepackage[normalem]{ulem}
\usepackage{enumerate}

\def\av#1{\langle#1\rangle}
\usepackage{color}
\usepackage{hyperref}

\def\av#1{\langle#1\rangle}
\usepackage{color}

\begin{document}
\title{On the decentralized navigation of multiple packages on 
transportation networks}

\author{Samuel M. da Silva}
\affiliation{Departamento de F\'{i}sica, Universidade Federal do Cear\'{a}, 60451-970 Fortaleza, Cear\'{a}, Brazil}

\author{Saulo D. S. Reis}
\affiliation{Departamento de F\'{i}sica, Universidade Federal do Cear\'{a}, 60451-970 Fortaleza, Cear\'{a}, Brazil}

\author{Asc\^anio D. Ara\'ujo}
\affiliation{Departamento de F\'{i}sica, Universidade Federal do Cear\'{a}, 60451-970 Fortaleza, Cear\'{a}, Brazil}

\author{Jos\'e S. Andrade, Jr.}
\affiliation{Departamento de F\'{i}sica, Universidade Federal do Cear\'{a}, 60451-970 Fortaleza, Cear\'{a}, Brazil}

\date{\today}

\begin{abstract}
We investigate by numerical simulation and finite-size analysis the 
impact of long-range shortcuts on a spatially embedded
transportation network. 
Our networks are built from
two-dimensional ($d=2$) square lattices to be improved by the 
addition of long-range shortcuts added with probability
$P(r_{ij})\sim r_{ij}^{-\alpha}$
[J. M. Kleinberg, Nature 406, 845 (2000)].
Considering those improved networks, we performed numerical
simulation of multiple discrete package navigation and found a limit 
for the amount of packages flowing through the network. Such limit 
is characterized by a critical probability of creating packages 
$p_{c}$, where above this value a transition to a congested state 
occurs. Moreover, $p_{c}$ is found to follow a power-law,
$p_{c}\sim L^{-\gamma}$, where $L$ is the network size. Our results 
indicate the presence of an optimal value of
$\alpha_{\rm min}\approx1.7$,
where the parameter 
$\gamma$ reaches its minimum value and the networks are more 
resilient to congestion for larger system sizes.
Interestingly, 
this value is close to the analytically found value of $\alpha$ for 
the optimal navigation of single packages in spatially embedded 
networks, where $\alpha_{\rm opt}=d$. Finally, we show that the 
power spectrum for the number of packages navigating the network at 
a given time step $t$, which is related with the divergence of the 
expected delivery time, follows a universal Lorentzian function, 
regardless the topological details of the networks.
\end{abstract}
\maketitle

\section{Introduction}

The navigation problem consists of sending a message, or some piece
of information, from a given source node to a target node of a
network~\cite{ref:newman2010,ref:barthelemy2011}. Taking this 
perspective into account, communication networks, the internet, 
the network of streets for public transportation, all share the same 
basic purpose: to deliver the desired packages as faster as possible 
to their destiny, while maintaining its functionality. From this 
point of view, the 1967 small-world experiment proposed by the 
American psychologist Stanley Milgram is a paradigmatic 
example~\cite{ref:milgram1967}. The algorithmic approach of the 
experiment performed later by Kleinberg~\cite{ref:kleinberg2000}, 
not only had shown that a navigation guided solely by decentralized 
algorithms is capable of accomplish the task, but also that the 
underlying dimension of the spatially embedded network can 
drastically affect the expected delivery time. In the present work, 
we investigate the impact of the underlying geography of the 
transportation network on the navigation of multiple packages where 
a load limit to the network nodes exists.

In many transportation networks of interest in science and 
technology, efficient transport of information packages, energy, or 
even people is thought in terms of avoiding congestion rather than 
minimizing expected delivery 
time~\cite{ref:arenas2001,ref:guimera2002a,ref:guimera2002b}. For 
instance, when using a crowdsourcing map application to navigate a
city, the driver usually sacrifices time travel, adopting a longer 
detour in order to avoid traffic jams. Consider now that, by the 
addition of new shortcuts, we aim to plan or improve an existing 
transportation network. As we shall show, to consider the underlying 
structure of such network, while adding shortcuts, plays an 
important role in the way the transportation occurs, allowing the 
increase of the number of packages traveling the network, without 
missing its functionality.

The framework of spatially embedded networks makes use of a regular
lattice of dimension $d$ with long-range connections randomly added
upon it. Generally, it considers the addition of long-range 
connections between two given nodes $i$ and $j$ with a probability 
decaying with their lattice distance
$r_{ij}$, $P(r_{ij})\sim r_{ij}^{-\alpha}$~\cite{ref:kleinberg2000}. 
The interplay between the underlying regular structure and a 
randomized long-range construction is a well-known recipe to mimic 
the so called small-world phenomenon~\cite{ref:watts1998}, where the 
typical distance $\ell$ between a pair of nodes grows slowly with 
the number of nodes $N$ of the network, 
$\ell\sim{\rm log}N$~\cite{ref:barthelemy1999}. However, for the 
case of spatially embedded networks, this is true only for 
$\alpha\leq d$~\cite{ref:kosmidis2008,ref:chen2016}. Remarkably, the 
small-world property can be accessed by a decentralized algorithm 
only when 
$\alpha=d$~\cite{ref:kleinberg2000,ref:carmi2009,ref:cartozo2009,
ref:hu2011a,ref:hu2011b}, a result that holds for 
fractals~\cite{ref:roberson2006,ref:rozenfeld2010}, for transport 
phenomena that obeys local conservation
laws~\citep{ref:oliveira2014,ref:sampaio2016}, nonlocal
percolation rules~\cite{ref:reis2012} and brain
networks~\cite{ref:gallos2012a,ref:gallos2012b}. Moreover, 
when a cost constrain is imposed on the addition of shortcuts to the 
underlying network, it has been found that the better conditions to 
navigation are attained when 
$\alpha=d+1$~\cite{ref:gli2010,ref:yang2010,ref:yli2010,
ref:gli2013}. It is argued that such conditions, with and without 
cost constrain, are optimal due to strong correlations between the 
underlying spatial network and
the long-range structure, 
allowing the package holder
to find the shortest paths at the small-world
network~\cite{ref:kleinberg2000}.
It is claimed that such compromise 
between local and long-range structures leads to an effective 
dimension higher than the dimension of the underlying local 
structure~\cite{ref:daqing2011,ref:boguna2009}.

The remaining of this manuscript is organized as follows. In
Section~\ref{sec:model} we describe the spatially embedded network 
model proposed by Kleinberg~\cite{ref:kleinberg2000}. The rules 
for node overload used in the present study in order to mimic 
the onset of congestion are also presented in Sec.~\ref{sec:model}. 
In Section~\ref{sec:results} the results of our simulations and 
numerical analysis are presented, where we study the behavior of 
the order parameter, the scale of the critical point and the 
divergence of the characteristic time. We leave further discussion 
and conclusions to Section~\ref{sec:conclusions}.

\section{Model formulation}\label{sec:model}

Using a simple and general model based on a decentralized algorithm, 
we study the effects of nonlocality assuming three simple 
ingredients~\cite{ref:arenas2001,ref:guimera2002b}. The first is a 
physical spatially embedded structure where the transportation 
process takes place, in other words, the transport network 
itself. Second, we assume that the channels through which the 
information flows have limited capacity. Finally, the 
information navigating this network is composed of discrete 
packages. Without lack of generality, the important characteristics 
of the problem are obtained by the analysis of the navigation and 
congestion of these discrete packages.

\begin{figure}
  \includegraphics[width=0.7\columnwidth]{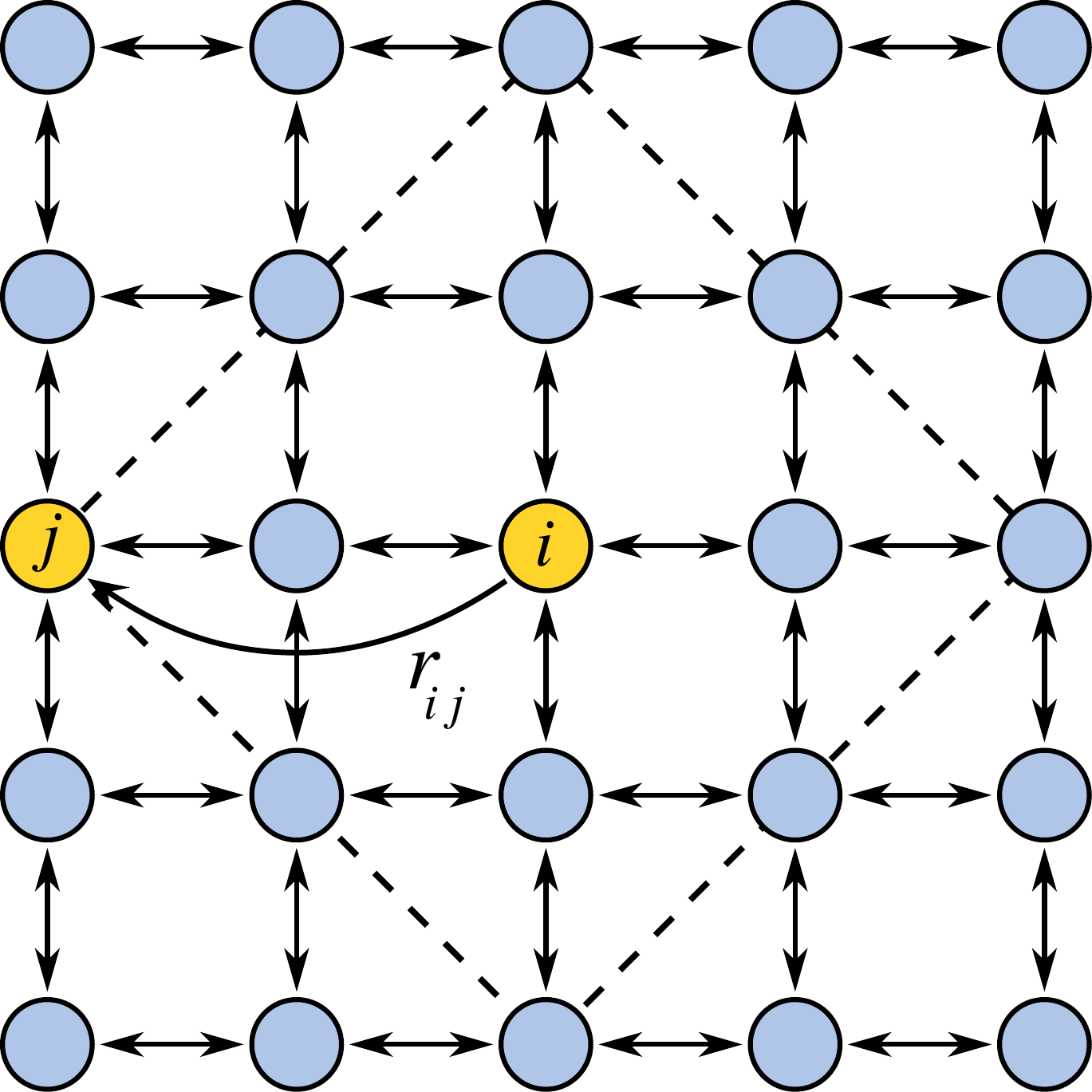}
  \caption{Kleinberg networks are built by adding long-range 	
  connections on regular lattices. For a given node $i$, a node $j$ 	
  is randomly chosen and an undirected link of length $r_{ij}$ is 
  added between them with probability
  $P(r_{ij})\sim r_{ij}^{-\alpha}$. By doing so, the node $j$ can be
  any node with a Manhattan distance of $r_{ij}$. The dashed lines 
  highlight the eight nodes separated from node $i$ by a lattice 
  distance $r=2$. We randomly choose the node $j$ from this set of 
  eight nodes.}
\label{fig:model}
\end{figure}

As shown in Fig.~\ref{fig:model}, the transportation network is 
embedded in a square lattice with $N=L\times L$ nodes upon which we 
add long-range connections. In this model, pairs of nodes $i$ and 
$j$ are chosen at random to receive one of those long-range 
connections with probability proportional to $r_{ij}^{-\alpha}$, 
where $r_{ij}$ is the Manhattan distance between nodes $i$ and $j$.
Since the number of nodes separeted by the lattice distance $r$ from 
a node $i$ in a $d$-dimensional lattice is proportional to $r^{d-1}$
(see Fig.~\ref{fig:model}), the probability $P(r_{ij})$ can be 
mapped into the density distribution function
$p(r)\sim r^{d-1-\alpha}$. After the distance $r$ is chosen 
following the distribution $p(r)$, we randomly choose node $j$ from 
the set of nodes separated from $i$ by the distance $r$. Clearly, 
the present model satisfies the small-world paradigm, i.e., it is 
rich in short-range connections, but has only a few long-range 
connections~\citep{ref:watts1998}. 

The package transportation algorithm is defined as follows. At 
first, we assign to the whole network a probability $p$ for the 
creation of information packages. After that, at each time step $t$, 
we draw a uniformly distributed random number in the interval 
ranging from 0 to 1 for each node $s$. If this number is smaller 
than $p$, a package of information is created at node $s$. Then, for 
each new package a target node $t$ is randomly assigned. In other to 
mimic real life situations, the nodes (e.g., the routers from the 
Internet) do not have information about the whole network topology. 
Therefore, an information holder node $a$ chooses from its set of 
neighbors, both short- and long-range, the neighbor node $b$ that is 
geographically closer to $t$ to send the package. Clearly, this 
algorithm has close relationship with the greedy algorithm proposed 
by Kleinberg utilized to study the problem of efficient navigation 
of one package of information in small-world
networks~\citep{ref:kleinberg2000}.
 
After the next potential holder is chosen, the package is 
transmitted or not from node $a$ to node $b$ through a channel 
(link) of quality $q_{ab}$~\cite{ref:arenas2001}. In real life 
situations, such quality influences the transmission probability, 
and it is expected to depend on the package load of the nodes it 
connects. Accordingly, the capacity $\kappa_a$ of node $a$ to 
receive a new package can be defined as
\begin{equation}
\kappa_a = \left\{
	\begin{array}{ll}
		1, & {\rm if~}n_a = 0;\\
		n_a^{-\xi}, & {\rm if~} n_a=1, 2, 3, ...;
	\end{array}\right.
\end{equation}
where $n_a$ is the number of packages at node $a$. Thus, one can 
define the channel quality as the geometric average 
$q_{ab}=\sqrt{\kappa_a\kappa_b}$~\citep{ref:arenas2001}. Then, we 
can rewrite channel quality as
\begin{equation}
q_{ab} = (n_{a}n_{b})^{-\xi/2}.
\end{equation}
Assuming $q_{ab}$ as the probability of node $a$ to deliver a 
package to node $b$, the average number $\langle n_{ab}\rangle$ of 
packages delivered from $a$ to $b$ per time unit will be 
proportional to $n_a^{1-\xi/2}n_b^{-\xi/2}$~\citep{ref:arenas2001}.
Note that the packages are uniformily created across 
the network, therefore, we can assume $n_a\sim n_b$. Consequently, 
\begin{equation}
\langle n_{ab}\rangle\sim n_{a}^{1-\xi}.
\end{equation}
When $\xi < 1$, the average number of delivered packages increases 
with the package load of nodes, and for this reason, the system is 
always found at a free-flow phase. On the other hand, if $\xi>1$, 
the average number of delivered packages decays fast, meaning that 
nodes fail to deliver the packages, causing the emergence of a 
congested phase. Along these lines, the present model has a clear 
phase transition from a free-flow phase to a congested one at 
$\xi=1$. In what follows, we focus our study at this critical value 
of $\xi$.

\section{Results and Discussion}\label{sec:results}

\subsection{Transition from free phases to congested phases}

According to the previous discussion, for $\xi=1$, we 
expect to have both free flow and congestion phases depending of the 
probability $p$. Therefore,
we start our analysis by defining and computing the order parameter 
$\eta$ given by
\begin{equation}
\eta(p) = \lim_{t\rightarrow\infty}\frac{1}{pL^2}\frac{\av{\Delta N}}{\Delta t},
\end{equation}
where $pL^2$ is the average number of packages created per time 
step, $\Delta N$ is the number of undelivered packages navigating 
the network at time windows of duration $\Delta t$. At stationary 
states ($t\rightarrow\infty$), the average $\av{\Delta N}$ is 
related with the rate of creating packages and the rate of 
delivering packages per time step, and therefore, it depends only on 
$p$ at large $t$, since $\Delta N$ will be proportional to the 
duration of the time windows
$\Delta t$~\citep{ref:arenas2001,ref:guimera2002b}.

\begin{figure}[!t]
  \includegraphics[width=0.8\columnwidth]{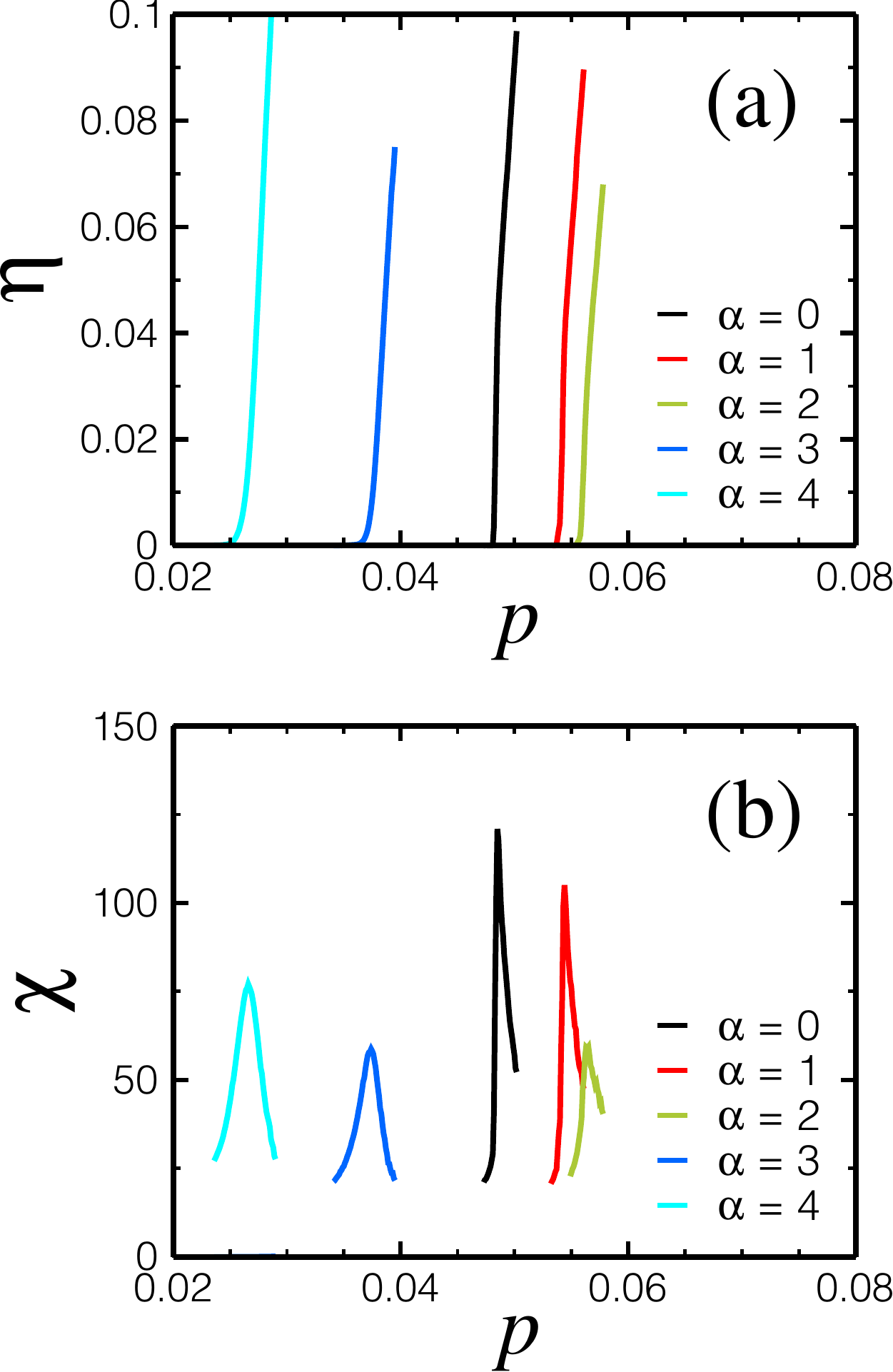}
  \caption{Transition to congested phases for different values of 
  $\alpha$. In (a) we show the order parameter $\eta$ as a function 
  of the probability of creating new packages $p$. When the network
  topology favors the decentralized delivery algorithm, the packages
  are delivered in a high rate than new packages are created,
  avoiding nodes to become overloaded, therefore preventing 
  congestion. This   is the case of $\alpha=0$, $1$, and $2$, 
  where the transition occurs for higher values of $p$. As depicted 
  in (b), the critical threshold $p_c$ marks a maximum value on the 
  response susceptibility $\chi$. Here, we use $L=128$ with $T=10^4$ 
  for each value of $\alpha$ and simulation time equal to $10^5$.}
\label{fig:transition}
\end{figure}

As depicted in Fig.~\ref{fig:transition}(a), the analysis of the 
order parameter $\eta$ allows us to define the critical probability 
of creating new packages $p_c$, characterizing a typical 
second-order phase transition between congested and free
phases~\citep{ref:arenas2001,ref:guimera2002a,ref:guimera2002b}. For 
small values of $p$, $\av{\Delta N}=0$ at the stationary state, 
resulting on free phases, meaning that the rate of creating packages 
is smaller than the rate of delivering packages. As $p$ increases, 
the rate of creating packages also increases, till it reaches the 
critical threshold $p_c$, where the rate of created packages is 
equal to the supported rate of delivering packages. Beyond this 
critical threshold, the value of $\av{\Delta N}$ increases with $t$, 
resulting on values for $\eta>0$, marking the existence of a 
congested phase.

Interestingly, $p_c$ presents a nontrivial behavior with $\alpha$ on 
Kleinberg networks, as depicted in Fig.~\ref{fig:transition}(a). 
Specifically, when $\alpha>0$, the critical threshold slightly 
increases ($\alpha=1$ and $2$), showing that at this range of 
$\alpha$ the networks are more resilient to the creation of 
information packages. However, for larger values of the exponent
$\alpha$, $\alpha=3$ and $4$, the critical threshold drastically 
decreases leading to networks more prone to congestion.

\subsection{Behavior of the critical threshold}

Next, we investigate the behavior of $p_c$. Second-order phase 
transitions are expected to be dominated by larger fluctuations on 
the order parameter close to their critical
threshold~\citep{ref:stanley1971}. Thus, in order to compute $p_c$ 
with precision, it is convenient to define a macroscopic 
susceptibility function $\chi(p)$ as
\begin{equation}
\chi(p) = \lim_{T\rightarrow\infty}T\sigma_{\eta}(T),
\label{eqn:sus}
\end{equation}
where $\sigma_{\eta}(T)$ is the standard deviation of the order 
parameter computed in many time windows of width $T$. Hence, in 
order to compute $\chi(p)$, it is necessary quite a long simulation 
time for one single realization.
In the context of critical phenomena, such function is sensitive to 
fluctuations of the order parameter, where it diverges as the 
control parameter approaches its critical 
value~\cite{ref:stanley1971}. As shown in 
Fig.~\ref{fig:transition}(b), $\chi(p)$ presents a maximum at 
nontrivial values of $p$ for different value of $\alpha$. 
Accordingly, we identify these values as
$p_c$~\citep{ref:arenas2001,ref:guimera2002b}.

\begin{figure}
  \includegraphics[width=0.85\columnwidth]{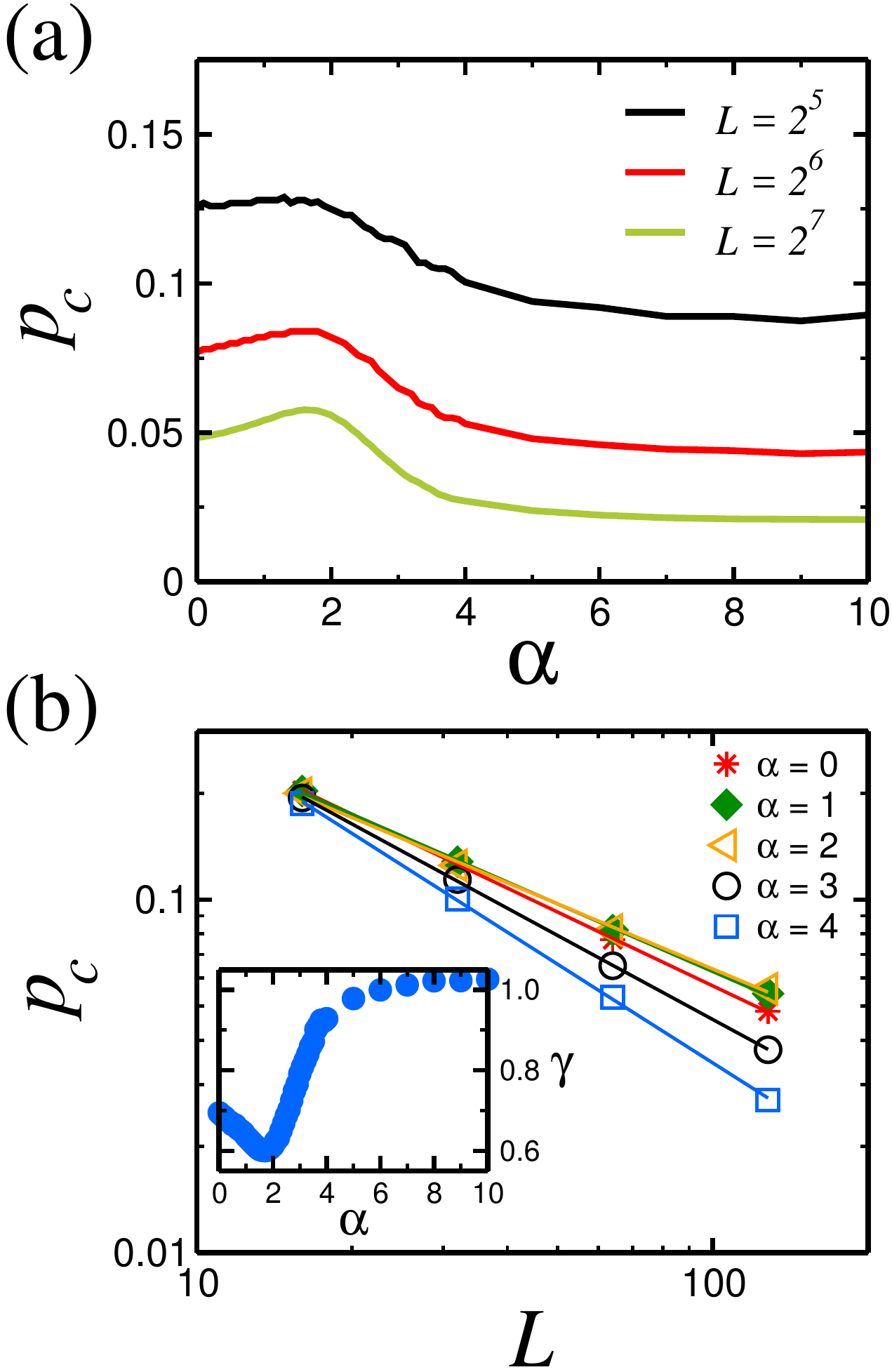}
  \caption{Behavior of the critical point $p_c$. In panel (a) we 
  show the critical probability $p_c$ as a function of the exponent 
  $\alpha$. As the system size $L$ increases, the packages take 
  longer to be delivered, causing the increase of $\Delta N$. As a 
  consequence of that, the transition to congested phases occurs for 
  even smaller values of $p_c$. However, independently of the value 
  of $L$, the critical probability always presents a maximum as 
  $\alpha$ approaches the value of $\alpha=2$. The critical 
  probabilities follow a power-law behavior, $p_c\sim L^{-\gamma}$, 
  as presented in the main plot of (b). The values of the exponent
  $\gamma$ resulting from a least-square fitting to the data are 
  presented in the inset, where a non-monotonic behavior can be
  clearly observed. Here, we use $L=2^5$ with $T=10^4$ for each 
  value of $\alpha$ and simulation time equal to $10^5$. }
\label{fig:criticalpoint}
\end{figure}

In order to determine how the exponent $\alpha$ affects the 
network's resilience to congestion, we study the dependence of the 
critical point $p_c$ with the linear size $L$ for different values 
of $\alpha$. Figure~\ref{fig:criticalpoint}(a) shows the estimated 
values of $p_c$ as a function of the exponent $\alpha$ for three 
different system sizes. In accordance with the results for $\eta(p)$ 
and $\chi(p)$, the critical point $p_c$ presents similar behavior 
with respect to $\alpha$. As presented in Fig.~\ref{fig:transition}, 
$p_c$ increases with $\alpha$ for $0<\alpha<2$. However, it 
decreases for $2<\alpha<4$, and finally saturates and reaches its 
lower values for $\alpha>4$.

The way in which $p_c$ scales with $L$, however, follows rather 
different behaviors, depending of the values of $\alpha$. As 
presented in the main plot of Fig.~\ref{fig:criticalpoint}(b), our 
results suggest that the critical point $p_c$ follows a scaling law 
with size $L$ following a power-law function, $p_c\sim L^{-\gamma}$. 
This power-law behavior provides an important piece of information 
about our model: as the system size increases, and consequently the 
distance between source nodes and target nodes also increases in 
average, more packages are navigating the network at a given time 
step. This makes larger networks more susceptible to congestion, 
since the expected number of packages occupying a node increases. 
For the case of mobility patterns in cities, this result agrees with 
the allometric relations between the city population and the total 
traffic delay~\cite{ref:louf2014} and traffic
accidents~\cite{ref:melo2014}. Moreover, the different values of the 
scaling exponent $\gamma$ suggest that for a given system size $L$, 
there are values of $\alpha$ that generate networks more robust to 
package transportation.

We extract more information about this scaling by performing 
extensive simulations for different values of $\alpha$ and very
long realizations for different system sizes. In each case, the
critical point $p_c$ is estimated by the computation of the 
susceptibility $\chi(p)$. The inset in 
Fig.~\ref{fig:criticalpoint}(b) shows the values of the exponent
$\gamma$ as a function of exponent $\alpha$. As one can see, as 
$\alpha$ increases, $\gamma$ slightly decreases from $\gamma=0.6(9)$ 
at $\alpha=0$, to close to $\gamma=0.6(1)$, when $\alpha$ approaches 
the dimension of the underlying network, $\alpha=2$. Here, we obtain 
the minimum value of $\gamma$ for $\alpha_{\rm min}=1.7$, where 
$\gamma=0.6(0)$. For $2<\alpha<4$, the value of the exponent 
$\gamma$ sharply increases. For the range of $4<\alpha<10$, the 
values of $\gamma$ increases very slowly. We found the values of 
$\gamma=0.9(3)$ for $\alpha=4$, and $\gamma=1.0(3)$ for $\alpha=10$.

The nonmonotomic behavior of $\gamma$ can be better understood by 
the analysis of the single-package navigation case. For this case, 
under decentralized algorithms, similar behavior was found in the 
same topology for the scaling of the expected delivery 
time~\cite{ref:kleinberg2000,ref:rozenfeld2010,ref:carmi2009,
ref:cartozo2009}. Precisely, the value of the exponent $\alpha$ that 
optimizes the delivery time of a package in two dimensions is
$\alpha_{\rm opt}=2$, where the expected delivery time scales 
logarithmic with $L$. For $\alpha$ different from the underlying 
network dimension $d$, the expected delivery time has a power-law behavior, 
$L^x$~\cite{ref:carmi2009,ref:cartozo2009}. In this situation, the 
value $\alpha=d$ is a transient point for which, 
$x = (d-\alpha)/(d+1-\alpha)$ for which $\alpha<d$, 
and $x = \alpha-d$ for $\alpha>d$. We believe that the different 
value found for the minimum, $\alpha_{\rm min}=1.7$, results from 
the small network sizes used due to the long simulation time 
necessary for the computation of $\chi(p)$. Hence, we conjecture 
that this optimal value, $\alpha_{\rm opt}=2$, would help in 
postponing the emergence of congestion for the case of multiple 
package navigation.

\begin{figure}[t!]
  \includegraphics[width=0.85\columnwidth]{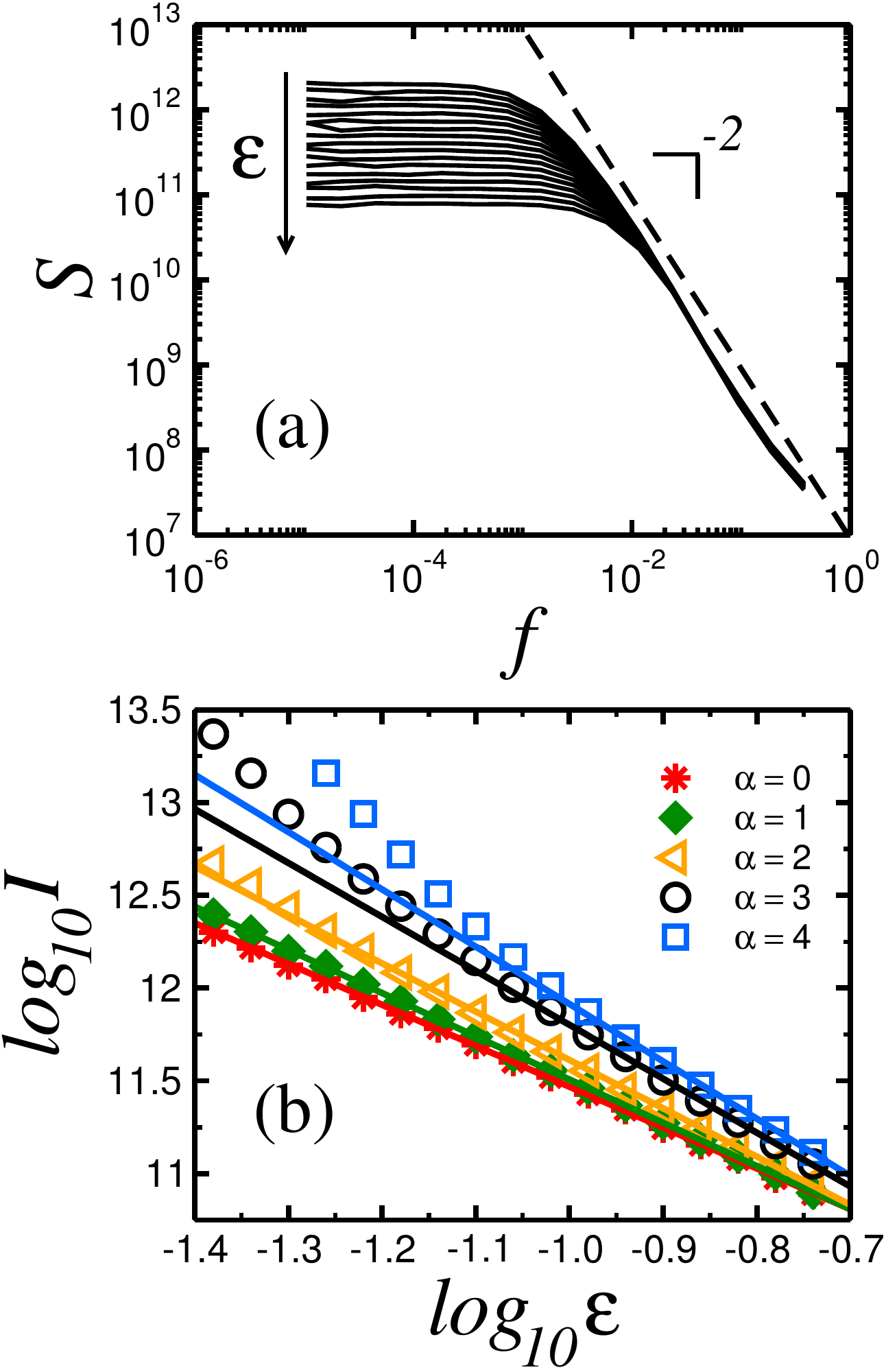}
  \caption{(a) Power spectrum of $S$ as a function of $f$ for 
  $\alpha=0$, and for the free phase, $p<p_c$. As one can 
  see, for all values of the rescaled control parameter 
  $\varepsilon$, $S$ has the form of a Lorentzian function given by
  Eq.~(\ref{eqn:lorentz1}). The values of $\varepsilon$ increase 
  from the top to the bottom. Therefore, the curves at the top are 
  closer to the critical value $p_c$. Provided the Lorentzian 
  behavior, it is expected a power-law decay of $S$ for larger 
  values of $f$, as shown by the dashed line with slope $-2$.
  (b) Intensity $I$ as a function of the order parameter
  $\varepsilon$ obtained from the non-linear curve fitting of
  $S$ using Eq.~(\ref{eqn:lorentz1}), for different values of 
  $\alpha$ at the free phase. The plot suggests that 
  $I\sim\varepsilon^{-\zeta}$. For $\alpha=3$ and $4$, the estimated 
  values of $I$ deviates from this behavior when the system 
  approaches the transition at the lower values of $\varepsilon$. 		
  The solid lines are the fitting results using 
  $I\sim \varepsilon^{-\zeta}$ at the ranges of $\varepsilon$ where 
  such power-law behavior holds. Each power spectrum is obtained 
  through an average of 100 realizations, with $L=32$ and 
  simulation time equal to $10^5$.} 
\label{fig:intensity_power_series}
\end{figure}
\begin{figure*}[t!]
  \includegraphics[width=2\columnwidth]{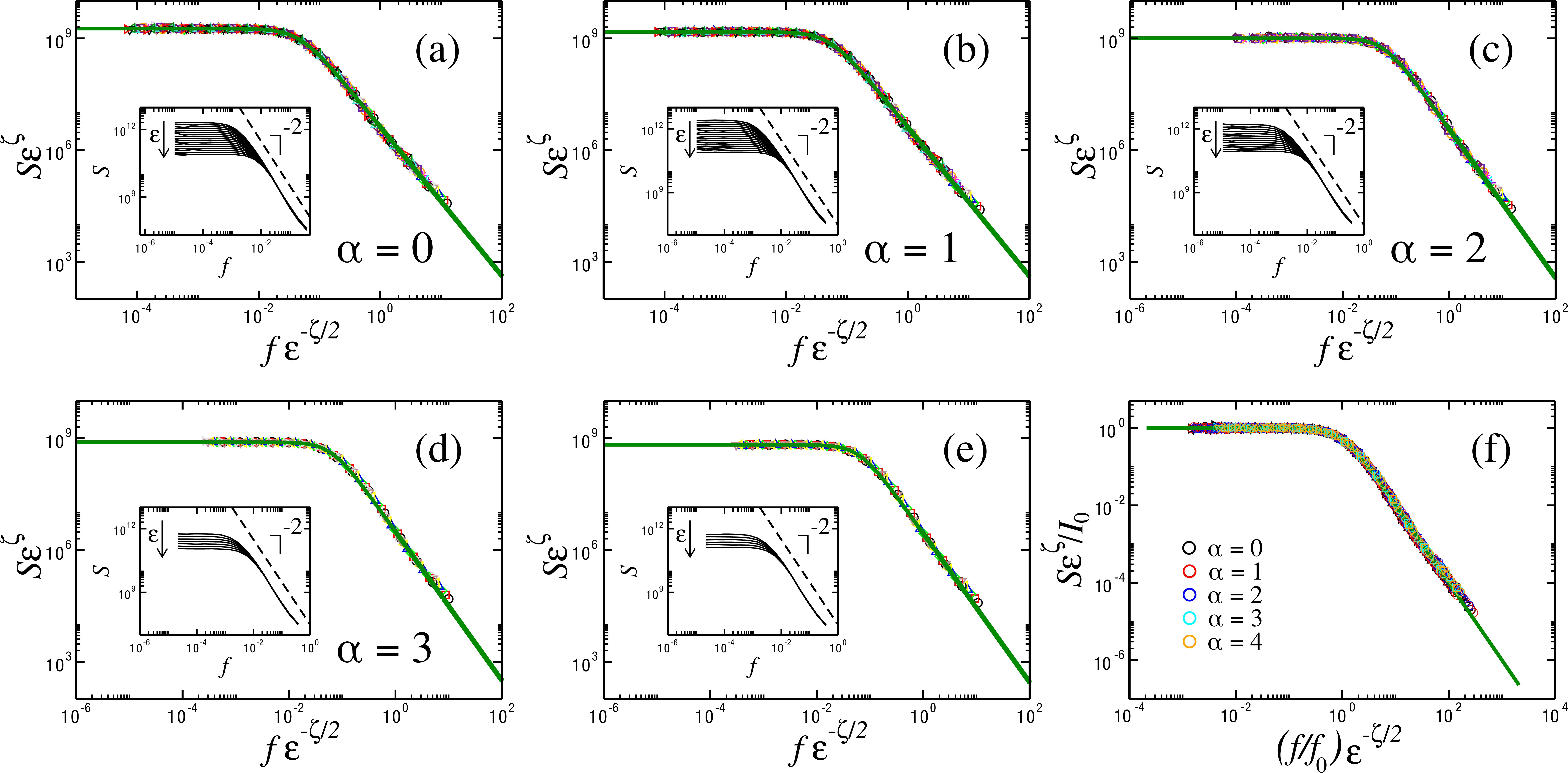}
  \caption{Data collapse of the power spectrum $S$ for different 
  values of the control parameter $\varepsilon$ and (a) 
  $\alpha=0$, (b) $\alpha=1$, (c) $\alpha=2$, (d) $\alpha=3$, and 
  (e) $\alpha=4$. We use $\zeta=2.2(0)$, $2.3(3)$, $2.6(0)$, 
  $2.9(1)$, and $3.0(9)$ for $\alpha = 0$, $1$, $2$, $3$, and $4$, 
  respectively. These values were obtained from a least-square 
  fitting to the data presented at 
  Fig.~\ref{fig:intensity_power_series}(b). The inset in each plot 
  is the original data for the power spectrum $S$ obtained from our 
  simulations. (f) When the plot axes are scaled using 
  $S\varepsilon^\zeta/I_{\rm 0}$ and 
  $\left(f/f_{\rm 0}\right)\varepsilon^{-\zeta/2}$, all data 
  presented at the panels (a)-(e) collapse into the universal 
  Lorentzian function $\mathcal{L}$. Each power spectrum is obtained 
  through an average of 100 realizations, with $L=32$ and 
  simulation time equal to $10^5$. } 
\label{fig:collapse}
\end{figure*}

\subsection{Divergence of expected delivery time}

Now, we focus our attention on the divergence of the expected 
delivery time. To do this, we analyze the behavior of the power 
spectrum of $N(t)$ defined as the Fourier transform
$S(f)=\mathcal{F}\{N(t)\}$, where $f$ is the frequency associated 
with package delivery. In Fig.~\ref{fig:intensity_power_series}(a), 
we show $S$ as a function of $f$ for different values of the 
rescaled control parameter $\varepsilon=|p-p_c|/p_c$ at the 
free phase, $p<p_c$, with $\alpha=0$. As depicted, 
the power spectrum $S$ has the form of a Lorentzian function given 
by:
\begin{equation}
S=\frac{I}{1+\left(\frac{f}{f_{\rm c}}\right)^2},
\label{eqn:lorentz1}
\end{equation}
where the intensity $I$ is the maximum value of $S$, and $f_{\rm c}$ 
is a characteristic frequency. The intensity parameter $I$ is 
associated with the the width of the Lorentzian function. On the 
other hand, the characteristic frequency $f_{\rm c}$ is closely 
related with the average time of delivering a 
package $\tau\sim 1/f_{\rm c}$. As one can see in
Fig.~\ref{fig:intensity_power_series}(a), as $\varepsilon$ 
decreases, the width of the plateau marking the value of $I$ 
also decreases, and at the limit $\varepsilon\rightarrow 0$ the 
power spectrum must scales as $1/f^2$. This is a signature of 
the divergence of the average expected delivery time, leading 
to $f_{\rm c}\rightarrow 0$ as $\varepsilon\rightarrow 0$, since
$\tau\sim\varepsilon^{-z}$~\cite{ref:guimera2002a}.
 
To better analyze the divergence of $\tau$ close to the transition, 
we compute the values of the exponent $z$ through the analysis of 
$I$ as a function of $\varepsilon$. As $\varepsilon\rightarrow 0$,
it is expected that $I\sim\varepsilon^{-\zeta}$.
Moreover, the exponent $z$, describing the divergence of the 
expected delivery time, must be related with $\zeta$ by the equation 
$z=\zeta/2$~\cite{ref:guimera2002a}. In 
Fig.~\ref{fig:intensity_power_series}(b), we show the values of $I$, 
collected from the nonlinear curve fitting of $S$ using
Eq.~\ref{eqn:lorentz1}, as a function of $\varepsilon$ in a log-log 
fashion. For $\alpha=0$, $1$, and $2$, we found 
$\zeta=2.2(0)$, $2.3(3)$, and $2.6(0)$, respectively. For $\alpha=3$ 
and $4$, the behavior of $I$ differs from such scaling as 
$\varepsilon\rightarrow 0$. Due to the 
Lorentzian profile, as $p$ approaches its critical value $p_c$, the 
width of the Lorentzian function diminishes, since $S\sim 1/f^2$ at this limit, 
resulting on an overestimation of $I$. Thus, in order to 
estimate $\zeta$, we use the range of $\varepsilon$ where the 
power-law behavior of $I$ holds. By doing this, we found 
$\zeta=2.9(1)$ and $3.0(9)$ for $\alpha=3$ and $4$, respectively.
Therefore, since $\tau\sim\varepsilon^{-\zeta/2}$, the monotonic 
increase of $\zeta$ with $\alpha$ leads to the conclusion that 
$\tau$ diverges faster for higher values of $\alpha$ as the system 
approaches the transition point. The results for $\zeta$ and the 
resilience to congestion with system size 
(see Fig.~\ref{fig:criticalpoint}) leads to a better compromise for 
transportation conditions attained when $\alpha\approx2$. Therefore,
based on previous results reported in the 
literature~\cite{ref:kleinberg2000,ref:kosmidis2008,
ref:carmi2009,ref:cartozo2009,ref:roberson2006,
ref:oliveira2014,ref:gli2010,ref:yang2010,ref:yli2010,ref:gli2013}, 
we conjecture that the optimal condition to the navigation of 
multiple packages on spatially embedded networks is obtained when
$\alpha=d$, where $d$ is the dimension of the underlying network.

Interestingly, the scaling relations $I=I_{\rm 0}\varepsilon^{-\zeta}$ and 
$f_{\rm c}=f_{\rm 0}\varepsilon^{\zeta/2}$ allow us to write
Eq.~\ref{eqn:lorentz1} in the more general form
\begin{equation}
S' = \varepsilon^{-\zeta}S(f\varepsilon^{\zeta/2}).
\label{eqn:lorentz2}
\end{equation}
As depicted in Figs.~\ref{fig:collapse}(a)-(e), all data for 
$S\varepsilon^\zeta$ as a function of $f\varepsilon^{-\zeta/2}$
collapse into a unique behavior for all values of $\alpha$. If one 
uses the values obtained for $I_{\rm 0}$ and $f_{\rm 0}$
(see Fig.~\ref{fig:collapse}(f)), the rescaling of the plot axes shows 
that, regardless of the topological details of the network defined 
by $\alpha$, all simulation data collapse into a universal curve 
described by the Lorentzian
\begin{equation}
\mathcal{L} = \frac{1}{1+F^2},
\label{eqn:lorentz3}
\end{equation}
where $F = (f/f_{\rm 0})\varepsilon^{-\zeta/2}$ and 
$\mathcal{L}=S'/I_{\rm 0}$. Since the addition of shortcuts change 
the effective dimension of the network~\cite{ref:daqing2011}, we 
believe that this Lorentzian behavior must hold for any network 
topology, such as the one-dimensional and two-dimensional lattices, 
as well as the Cayley tree~\cite{ref:guimera2002a}.

\section{Conclusions}\label{sec:conclusions}
In order to reveal the role of network topology on the transport of 
information packages, we studied the dynamics of a simple and 
general model of transportation networks on spatially embedded
networks. By assuming additional long-range connections added to a 
two dimensional square lattice following a power-law distribution
$P(r_{ij})\sim r_{ij}^{-\alpha}$, we found a limit for the total 
amount of information packages flowing through the network 
characterized by a critical probability of creating new packages 
$p_c$. Our results show that $p_c$ is described by a power-law of the 
network linear size, $p_c\sim L^{-\gamma}$, as it has been found in 
other topologies~\cite{ref:guimera2002b}. Remarkably, 
due to the characteristics of the network model used, $\gamma$ 
presents a nontrivial dependence with the topological parameter 
$\alpha$. Specifically, $\gamma$ has a minimum for
$\alpha_{\rm min}=1.7$, meaning that, in this case, the network is 
more resilient to the creation of new information packages. Since 
the transportation algorithm makes use only of local knowledge of 
the network geography, this robustness condition 
coincides with the small-world regime of the
network~\cite{ref:kosmidis2008}. Moreover, it is close to the 
optimal navigation condition for a single package on two dimensions, 
$\alpha_{\rm opt}=2$. Therefore, the spatial peculiarities of the 
network play a major role on the robustness of multiple package 
transportation. In other words, when optimizing the 
resilience of information exchange on transportation networks, one 
should take into account not only the protocol adopted, but also 
look for hints provided by the analysis of its geographical
properties. These results leads us to conjecture that the optimal 
navigation of multiple packages in spatially embedded networks is 
attained when $\alpha_{\rm opt}=d$, in the same way as in the 
optimal navigation of single packages~\cite{ref:kleinberg2000,
ref:kosmidis2008,ref:carmi2009,ref:cartozo2009,ref:roberson2006,
ref:oliveira2014,ref:gli2010,ref:yang2010,ref:yli2010,ref:gli2013}.

In addition, beyond the robustness of transportation networks, we 
studied its critical behavior by analyzing the power spectrum of the 
total number of packages as a function of time. When $p<p_c$, these 
spectra are described by a Lorentzian function, and saturate at a 
characteristic value $I$. In contrast, when $p$ approaches $p_c$, 
the power spectra present a power-law behavior with exponent $-2$. 
Considering the characteristic saturation value $I$ and its power-law 
behavior described by the exponent $\zeta$, we were able to show 
that the power spectra collapse into a universal Lorentzian 
function, regardless the topological details of the networks and 
their embedding dimension. These power spectra provide the 
divergence of relevant quantities for practical purposes, as the 
average time of delivering a package and the number of packages 
navigating the network. We expect that the modeling approach and 
results presented here could be useful in further studies on the 
multiple package navigation problem in different network topologies, 
which might lead to significant improvements on the ever-present 
package delivering processes occurring in real transportation 
networks.

\section*{Acknowledgments}

We thank the Brazilian agencies CAPES, CNPq, FUNCAP, FINEP and the 
National Institute of Science and Technology for Complex System in 
Brazil (INCT-SC).

\end{document}